\documentclass[twocolumn,twoside]{IEEEtran}

\ifCLASSINFOpdf
  
\else
 
\fi

\ifCLASSOPTIONcompsoc
    \usepackage[caption=false, font=normalsize, labelfont=sf, textfont=sf]{subfig}
\else
    \usepackage[caption=false, font=normalsize]{subfig}
\fi
\usepackage{lipsum}% 

\usepackage{balance}
\usepackage{multicol}   % for equalize of last page columns
\usepackage{cite}
\usepackage{gensymb}
\usepackage{multirow}
\usepackage{graphics}  
\usepackage{epsfig} 
\usepackage{graphicx}
\usepackage{epstopdf}
\usepackage{textcomp}
\usepackage{amsmath}
\usepackage{mathtools}
\usepackage{filecontents}
\usepackage{lipsum,color}
\usepackage{amsfonts}

\usepackage{mathrsfs}
\usepackage{textpos}

\begin{document}
\title{Ergodic Capacity of Triple-Hop All-Optical Amplify-and-Forward Relaying over Free-Space Optical Channels}

\author{\IEEEauthorblockN{Mohsen~Naseri,~Mohammad~Taghi~Dabiri~and~Seyed~Mohammad~Sajad~Sadough}\\
	\IEEEauthorblockA{Department of Electrical Engineering, Shahid Beheshti University G. C., Tehran, Iran \\
		Email: \{m\_naseri, m\_dabiri, s\_sadough\}@sbu.ac.ir}}
\maketitle

%\author{Mohsen~Naseri,~Mohammad~Taghi~Dabiri~and~Seyed~Mohammad~Sajad~Sadough
%\author{M.T.~Dabiri~and~S.M.S.~Sadough
%\thanks{M. T. Dabiri and S. M. S. Sadough are with the Department of Electrical Engineering, Shahid Beheshti University G. C., 1983963113, Tehran, Iran (e-mail: \{m$\_$dabiri, s$\_$sadough\}@sbu.ac.ir).}
%}    

% The paper headers

%\markboth{$1^{\rm st}$ West Asian Colloquium on Optical Wireless Communications (WACOWC2018)}
%{Ergodic Capacity of Triple-Hop All-Optical Amplify and Forward Relaying over Log-Normal FSO Channels.}

% make the title area
\maketitle

%%%%%%%%%%%%%%%%%%%%%%%%%%%%%%%%%%%%%%%%%%%%%%%%%%%%%%%%%%
%%%%%%%%%%%%%%%%%%%%%%%%%%%%%%%%%%%%%%%%%%%%%%%%%%%%%%%%%%
\begin{abstract}
%%%%%%%%%%%%%%%%%%%%%%%%%%%%%%%%%%%%%%%%%%%%%%%%%%%%%%%%%%
%%%%%%%%%%%%%%%%%%%%%%%%%%%%%%%%%%%%%%%%%%%%%%%%%%%%%%%%%%
In this paper, we propose a comprehensive  research over triple hop all-optical relaying free-space optical (FSO) systems in the presence of all main noise sources including background, thermal and  amplified spontaneous emission (ASE) noise and by considering the effect of the optical degree-of-freedom (DoF).
Using full CSI relaying, we derive the exact expressions for the noise variance at the destination.
Then, in order to simplify the analytical expressions of full CSI relaying, we also propose and investigate the validity of different approximations over noise variance at the destination.
Finally, we evaluate the the performance of considered triple-hop all-optical relaying FSO system in term of ergodic capacity.

\end{abstract}
%%%%%%%%%%%%%%%%%%%%%%%%
\begin{IEEEkeywords}
 Free-space optics (FSO), atmospheric turbulence, all optical relay, log-normal distribution.

\end{IEEEkeywords}

\IEEEpeerreviewmaketitle

%\begin{textblock*}{3cm}(5.7cm,-12.2cm)
%	\makebox[0.1\columnwidth]{$1^{\rm st}$ West Asian Colloquium on Optical Wireless Communications (WACOWC2018)}
%\end{textblock*}
%%%%%%%%%%%%%%%%%%%%%%%%%%%%
%%%%%%%%%%%%%%%%%%%%%%%%%%%%
%%%%%%%%%%%%%%%%%%%%%%%%%%%%
%%%%%%%%%%%%%%%%%%%%%%%%%%%%
%%%%%%%%%%%%%%%%%%%%%%%%%%%%%%%%%%%%%%%%%%%%%%%%%%%%%%%%%%%%
%%%%%%%%%%%%%%%%%%%%%%%%%%%%%%%%%%%%%%%%%%%%%%%%%%%%%%%%%%%%
\section{Introduction}
\label{I}
%%%%%%%%%%%%%%%%%%%%%%%%%%%%%%%%%%%%%%%%%%%%%%%%%%%%%%%%%%%%
%%%%%%%%%%%%%%%%%%%%%%%%%%%%%%%%%%%%%%%%%%%%%%%%%%%%%%%%%%%%
The rapidly increasing need of very high data rate communications  made free space optical (FSO) communication very popular during the past few years.
FSO communication offers a great potential due to the ease of deployment, unregulated frequency band and the enormous available bandwidth \cite{ghassemlooy2012optical}.
FSO communication has observed great scientific developments and engineering improvements that make it usable for various commercial, industrial,  military, environmental, and sensor network applications \cite{Dr_khalighi_survay,dabiri2017performance}. %and they are now widely available in the market, and an FSO link can be set up quickly in several minutes to hours
%A major effect of the turbulence is the scintillation, which causes intensity fluctuations at the received signal. This significantly decreases the FSO link performance.
%In addition, for short range links, such high-bandwidth technology has been proposed as a viable alternative for high-speed “last-mile” connectivity, the backbone segments of mobile infrastructures \cite{rajakumar2008interference,smadi2009free}. 
However, atmospheric turbulence-induced fading, geometrical spread and channel loss, which are distance-dependent, significantly decrease the FSO link performance \cite{safari2012multi,anees2015performance,dabiri2017fso}. %soleimani2016generalized,douik2016hybrid,ansari2013performance
To overcome this limitation, relaying techniques to divide a long link into multiple short links, has been advocated for FSO transceivers as a mean to cover large distances and improve the performance of FSO systems \cite{boluda2015impact}. %dahrouj2015cost,

Research studies on relay-assisted FSO systems was initially focused on electrical relaying where electrical-optical and optical-electrical conversions was performed at  relays \cite{safari2008relay,kashani2013optimal,boluda2016miso}. However, it is well known that the efficiency of electrical relaying is limited due to the requirement of high-speed electrons and optoelectronic hardware. 
Moreover, by developing tracking systems using acousto-optic devices that especially applicable to the fast moving unmanned aerial vehicles (UAVs) for compensating the pointing losses, UAV-based FSO communication systems are employed for high data transmission between moving UAVs \cite{dabiriJSAC}.
However, due to the power, size and weight constraints of UAV systems, in practice, all-optical relaying is preferred to relying optical signals between UAVs \cite{vu2018performance}. 
To avoid the inherent complexities associated with electrical-optical and optical-electrical conversions in electrical relays, a new setup of relay-assisted FSO system referred to as all-optical relaying was recently proposed in the context of FSO communication where all processing are done in the optical domain \cite{yang2014performance,trinh2015alll,trinh2015all,nor2016comparison,nor2015performance,nor2017experimental,kazemlou2011all,bayaki2012,SAFARI_ALL_OPTIC_2012,IET2014,dabiri2017performance2}.
In particular, in \cite{kazemlou2011all}, optical amplify and optical regenerate techniques were developed and bit error rate (BER) performance was investigated by using Monte-Carlo (M-C) simulations. However, the latter work does not consider the effect of amplified spontaneous emission (ASE) noise. In \cite{bayaki2012}, the outage probability for a dual-hop FSO relay system was investigated in which ASE noise of optical amplifier was considered.
However, in \cite{bayaki2012}, the authors have not considered the effect of the optical degree-of-freedom (DoF).
In \cite{SAFARI_ALL_OPTIC_2012} and \cite{IET2014}, the outage probability of a dual-hop and parallel all-optical relaying was studied in which the effects of both ASE noise and optical DoF was considered.
However, performance analysis performed in \cite{SAFARI_ALL_OPTIC_2012,IET2014}, is based on the assumption that the effects of the background and thermal noises are negligible compared to the effect of the ASE noise.
Recently, in \cite{dabiri2017performance2} a comprehensive signal model for a more precise analysis of dual-hop all-optical relaying FSO systems is studied in the presence of all main noise sources. In \cite{dabiri2017performance2} the authors shown that due to the presence of background and ASE noise, the optimal location of the relay node is not placed at half distance of transceiver.

In this paper, we extend the results of \cite{dabiri2017performance2} and investigate the performance of triple-hop all-optical FSO  system relaying.
Particularly, using full CSI\footnote{Notice that this is a practical assumption due to the slow fading property of FSO links  \cite{dabiri2017generalized,dabiri2017glrt,WCL2018}} relaying, we derive the exact expressions for the noise variance at the destination in the presence of all main source noises including background, thermal and ASE noise by taking into account the presence of DoF.
Then, in order to simplify the analytical expressions of full CSI relaying, we also propose and investigate the validity of different approximations over noise variance at the destination.
The motivation of this paper is answering to this question: what is the relationship between relay locations and noise sources.
For this aim numerical results are plotted in term of ergodic capacity versus different values of source-to-relay, relay-to-relay and relay-to-destination.
Simulation results revealed that the optimal relay locations strongly depend on noise sources.

The rest of this paper is organized as follows.
In Section II, we describe the triple-hop FSO system model along with our main assumptions. 
In this Section, we derive the noise variance at the destination. we propose some approximations over noise variance at the destination.
In Section III, we study the performance of considered system in term of ergodic capacity throw numerical results. 
Finally in Section IV, we draw our conclusions.

%%%%%%%%%%%%%%%%%%%%%%%%%%%%%%%%%
%%%%%%%%%%%%%%%%%%%%%%%%%%%%%%%%%
\section{System Model}
%%%%%%%%%%%%%%%%%%%%%%%%%%%%%%%%%
%%%%%%%%%%%%%%%%%%%%%%%%%%%%%%%%%
%%%%%%%%%%%%%%%%%%%%%%%%%%%%%%%
%%%%%%%%%%%%%%%%%%%%%%%%%%%%%%%
\begin{figure*}
	\begin{center}
		\includegraphics[width=5 in ]{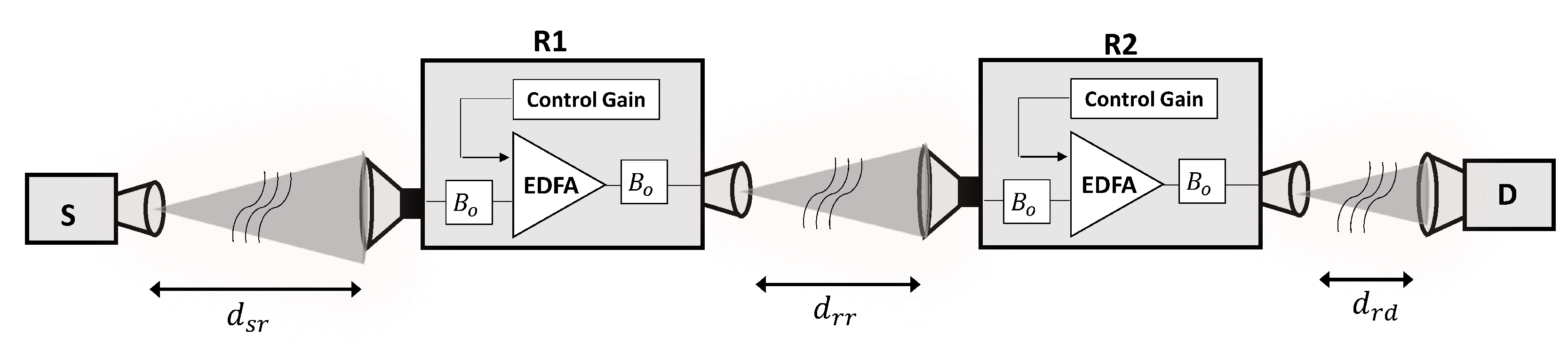}
		\caption{ All-optical triple-hop FSO system model.}
		\label{system_Model}
	\end{center}
\end{figure*}
%%%%%%%%%%%%%%%%%%%%%%%%%%%%
In this paper, we consider a relay-assisted intensity modulation-direct detection (IM-DD) FSO system, where the information at the source is transmitted to the destination through two all-optical AF relay network, i.e., the process of reception, amplification and retransmission are done in the optical domain.
The source-to-destination (SD) link length is $d_{sd}=d_{sr}+d_{rr}+d_{rd}$, where $d_{sr}$, $d_{rr}$ and $d_{rd}$ are source-to-relay (SR), relay-to-relay (RR) and relay-to-destination (RD) link length.
As depicted in Fig. \ref{system_Model}, the considered system model includes the following main parts: transmitter, turbulence channel, all-optical relays and receiver. In the sequel, we analyze these four parts with more details to obtain the mathematical models for the signals and noises at the destination in the context of all-optical relaying system.
%%%%%%%%%%%%%%%%%%%%%%%%%%%%%
\subsubsection{Transmitter}
%%%%%%%%%%%%%%%%%%%%%%%%%%%%%
%The laser is chosen by its central wavelength, average power, and beam divergence angle.
It is assumed that the transmitter is equipped with a laser diode operating at wavelength $\lambda=$ 1550 nm. Notice that the 1550 nm band is compatible with erbium-doped fiber amplifier (EDFA) technology, which is useful for long range FSO links to improve their performance \cite{razavi_2005}.
The mean photon counts of the transmitted pulses is $m_t = P_t T_s / h_p\nu$, where $P_t$, $T_s$, $\nu$ and $h_p$ are the transmitted power, chip duration, optical frequency and Plank's constant, respectively.
For convenience, we use the notations $m_s$, $m_{r1}$ and $m_{r2}$ to denote the mean of transmitted photon counts at source, first and second relay, respectively.
%%%%%%%%%%%%%%%%%%%%%%%%%%%%%%%%
\subsubsection{FSO Channel}
%%%%%%%%%%%%%%%%%%%%%%%%%%%%%%%%
We assume that the transmitted power is attenuated by two factors: the path loss $h_l$ and the atmospheric turbulence $h_a$. 
Hence, the channel gain $h$ can be formulated as $h=h_l h_a$.
The path loss $h_l$ is a function of the optical wavelength $\lambda$, divergence angle $\theta_{\rm div}$ and link length $d$, and can be expressed as \cite{loss_1}
\begin{align}
\label{jh1}
h_l = \frac{A_a}{\left(\theta_{\rm div} d/2\right)^2} h_l'~,
\end{align}
where $A_a$ is the receiver aperture area and $h_l'$ is the atmospheric attenuation which is modeled by the exponential Beers-Lambert Law as
$h_l'=\exp\left(-d\xi\right)$ and $\xi$ is the scattering coefficient \cite{loss_1}. 
For clear and foggy weather conditions, the scattering coefficient is a function of the visibility $\mathcal{V}$, and can be written as \cite{loss_1} 
\begin{align}
\label{st1}
\xi = \frac{3.91}{\mathcal{V}} \left(  \frac{\lambda}{550}\right)^{-q},
\end{align}
where $q$ is equal to 1.6, 1.3 and 0.585$\mathcal{V}^{1/3}$ for $\mathcal{V}>50$ Km, 50 Km$>\mathcal{V}>6$ Km and $\mathcal{V}<6$ Km, respectively.

We model the turbulence-induced fading $h_a$ by a log-normal distribution, i.e., $\ln\mathcal{N} (h_a;\mu_{l,h_a}, \sigma_{l,h_a}^2)$, which is emerged as a useful model for a weak atmospheric turbulence regime \cite{jurado2007efficient}.
The probability density function (PDF) of the log-normal model is given by
\begin{align}
\label{ch1}
f_{h_a}(h_a)=\frac{1}{h_a\sigma_{l,h_a}\sqrt{2\pi}} \exp\left( -\frac{\left(\ln h_a-\mu_{l,h_a}\right)^2}{2\sigma^2_{l,h_a}}\right).
\end{align}
To ensure that fading does not attenuate or amplify the average power, we normalize the fading coefficients such that $\mu_{h_a}=1$. Doing so requires the choice of $\mu_{l,h_a}=-\sigma^2_{l,h_a}/2$. Notice that $\sigma^2_{l,h_a}\simeq \sigma^2_{R}/4$ where $ \sigma^2_{R}$ is the Rytov variance and can be measured directly from atmospheric parameters as \cite{ghassemlooy2012optical}
\begin{align}
\label{ch2}
\sigma^2_{R} = 1.23 C_n^2 \kappa^{7/6} d^{11/6},
\end{align}
where $\kappa=2\pi/\lambda$ and $C_n^2$ is the refractive index structure constant.
In this paper, we assume that the transceivers are perfectly aligned.

Since, the channel gain is modeled as $h=h_l h_a$, hence, $h$ is distributed according to the log-normal distribution $\ln\mathcal{N} (h;\mu_{l,h}, \sigma_{l,h}^2)$ where $\mu_{l,h}=-\sigma^2_{R}/8+\ln(h_l)$ and $\sigma_{l,h}^2=\sigma^2_{R}/4$.
For convenience, we use the notations $h_{sr}$, $h_{rr}$  and $h_{rd}$ to denote the SR, RR and RD channel gains, respectively.
%%%%%%%%%%%%%%%%%%%%%%%%%%%%%%
\subsubsection{EDFA-Based Relays}
%%%%%%%%%%%%%%%%%%%%%%%%%%%%%%
At the relay nodes, the incoming optical light is guided into the optical fiber by a converging lens. 
However, in addition to the desired optical signal, the undesired background light due to the scattered sunlight is also collected by the receiver. 
The collected optical light is then amplified by an EDFA to compensate the channel loss. 
An EDFA is characterized by the noise factor $K=n_{sp}(G-1)$, the number of spontaneous mode\footnote{The number of spontaneous mode $D$ is the product of spatial and temporal modes and is obtained as $D = B_0 D_p T_s$, where $B_0$ is the bandwidth of optical filter that follows the preamplifier in Hz and $D_p$ is number of polarization modes \cite{EFA_book}.} $D$, and gain $G$ where, $n_{sp}$ is the amplifier spontaneous emission factor.
Assuming that the ASE noise is modeled by an additive noise (similar to the background noise), the photon count $n$, at the output of first relay is Laguerre distributed as \cite{EFA_book,razavi_2005}
\begin{align}
\label{j2d3}
f_{n,r_1}(n) = Lag\left\{ n, G_1 m_s h_{sr}, G_1 (m_{br_1} + n_{sp}) , D \right\},
\end{align}
where $m_{br_1}=P_{br_1}T_s/h_p\nu$ and $P_{br_1}$ is the filtered background power at first relay. 
The distribution of Laguerre (also known as the non-central negative-binomial) is given in \cite[Eq. (5)]{razavi_2005}. Moreover, the mean and the variance of $Lag\{n,a,b,D\}$ are respectively derived in \cite[Eq. (7)]{razavi_2005} as
\begin{align}
\label{lag1}
&m_{{\rm Lag}}(a, b, D) = a + Db, \nonumber \\
{\it Var}_{{\rm Lag}}&(a, b, D) = a+D(b+b^2)+2ab.
\end{align} 

Similar to the approach followed by most of existing work in the context of dual-hop FSO communications \cite{bayaki2012,IET2014,SAFARI_ALL_OPTIC_2012}, in this paper,  full CSI relaying (variable gain) is considered.
Note that, for FSO channels,  CSI can be easily obtained over a sequence of received bits \cite{dabiri2017generalized,dabiri2017glrt}.
In full CSI relaying, the instantaneous CSI of the SR and RR links are assumed to be known at the relays. First and second relays set the amplifier gains based on the instantaneous fading of the SR and RR links $h_{sr}$ and $h_{rr}$, respectively, in such a way that the relay output powers $P_{r_1}$ and $P_{r_2}$, are kept constant at all time-slots.
Due to the high gain of EDFAs (i.e., $G\gg 1$), in this paper, we approximate the noise factor as $K\simeq n_{sp}G$.
The amplifier gains of both first and second relays are calculated as \cite{SAFARI_ALL_OPTIC_2012}
\begin{align}
\label{fcsi}
G_1 =& \frac{m_{r_1}}{m_s h_{sr} + D(m_{br_1}+n_{sp})}, \\
\label{djh}
G_2 =& \frac{m_{r_2}}{m_{r_1} h_{rr} + D(m_{br_2}+n_{sp})},
\end{align}
where $m_{r_1}=P_{r_1} T_s / h_p\nu$ and $m_{r_2}=P_{r_2} T_s / h_p\nu$ are the average transmitted photons by first and second relays, respectively.
According to \eqref{j2d3}, the photon count $n$, at the output of second relay is distributed as
\begin{align}
\label{42d3}
f_{n,r_2}(n) = Lag\Big\{ &n, G_1G_2 m_s h_{sr} h_{rr}, G_1G_2 (m_{br_1} + n_{sp})h_{rr} \nonumber \\
&+G_2 (m_{br_2} + n_{sp}) , D \Big\},
\end{align}
where $m_{br_2}=P_{br_2}T_s/h_p\nu$ and $P_{br_2}$ is the filtered background power at second relay.
%%%%%%%%%%%%%%%%%%%%%%%%%%%
\subsubsection{Signal Model at Destination} 
%%%%%%%%%%%%%%%%%%%%%%%%%%%
The amplified optical signal is forwarded to the destination via a transmitting lens. 
At destination, the received optical signal, including the background radiation, is collected and focused on the surface of a photo-detector.
Based on \eqref{42d3}, the photon count $n$, at the input of photo-detector has also a Laguerre distribution as 
\begin{align}
\label{43d3}
f_{n,d}(n) =& Lag\Big\{ n, G_1G_2 m_s h_{sr} h_{rr}h_{rd}, G_1G_2 (m_{br_1}+ n_{sp})  \nonumber \\
& \times h_{rr}h_{rd} + G_2 (m_{br_2} + n_{sp})h_{rd} + m_{bd} , D \Big\},
\end{align} 
where $m_{bd}=P_{bd}T_s/h_p\nu$ and $P_{bd}$ is the filtered background power at destination.
At the destination, the collected optical signal is converted to an electrical one by a photo-detector (PD).
The Laguerre distribution has the property that it preserves its form after photo-detection and $Lag(n,a,b,D)$ is converted to $Lag\{n,\eta a, \eta b, D\}$ by a photo-detector with quantum efficiency $\eta$ \cite{razavi_2005}. 
Hence, the photon count distribution at destination, without considering the effect of thermal noise, is derived as 
\begin{align}
\label{laf5}
f'_{n,d}(n) =& Lag\Big\{ n, \eta G_1G_2 m_s h_{sr} h_{rr}h_{rd}, \eta G_1G_2 (m_{br_1} + n_{sp})  \nonumber \\
& \times h_{rr}h_{rd} + \eta G_2 (m_{br_2} + n_{sp})h_{rd} + \eta m_{bd} , D \Big\}.
\end{align} 
By substituting \eqref{laf5} in \eqref{lag1} and without considering the effect of thermal noise after photo-detection process, the mean and variance of photon count at  destination are respectively derived as 
\begin{align}
\label{lmw1}
 m'_d   &=  \eta G_1G_2 m_s h_{sr} h_{rr}h_{rd} \nonumber \\
 &~~~ + \eta D G_1G_2 (m_{br_1} + n_{sp}) h_{rr}h_{rd} \nonumber \\
 &~~~+ \eta D G_2 (m_{br_2} + n_{sp})h_{rd} +\eta D m_{bd},
 \end{align}
 and
% \begin{align}
% \label{lmw2}
%\sigma'^2_d=& \eta G_1G_2 m_s h_{sr} h_{rr}h_{rd} \nonumber \\
%5& + \eta D G_1G_2 (m_{br_1} + n_{sp})h_{rr}h_{rd} \nonumber \\
%&+ \eta D G_2 (m_{br_2} + n_{sp})h_{rd} + \eta D m_{bd} \nonumber \\
%%
%&+ D\big(\eta G_1G_2 (m_{br_1} + n_{sp})h_{rr}h_{rd}\big)^2    \nonumber \\
%&+ D\big(\eta  G_2 (m_{br_2} + n_{sp})h_{rd}\big)^2 + D\big(\eta m_{bd}\big)^2 \nonumber \\
%%
%&+ 2\eta^2 D G_1G_2^2 (m_{br_1} + n_{sp}) (m_{br_2} + n_{sp})h_{rr}h_{rd}^2 \nonumber \\
%&+ 2\eta^2 D G_1G_2 (m_{br_1} + n_{sp})m_{bd}h_{rr}h_{rd} \nonumber \\
%&+ 2\eta^2 D G_2 (m_{br_2} + n_{sp})m_{bd}h_{rd} \nonumber \\
%%
%&+ 2\eta^2 G_1^2G_2^2 m_s(m_{br_1} + n_{sp}) h_{sr} h_{rr}^2h_{rd}^2 \nonumber \\
%&+ 2\eta^2 G_1  G_2^2 m_s(m_{br_2} + n_{sp}) h_{sr} h_{rr}  h_{rd}^2     \nonumber \\
%&+ 2\eta^2 G_1  G_2   m_s m_{bd}             h_{sr} h_{rr}h_{rd}. 
%\end{align}
 \begin{align}
\label{lmw2}
\sigma'^2_d=& \eta G_1G_2 m_s\big(1+ 2\eta m_{bd}\big) h_{sr} h_{rr}h_{rd}  \nonumber \\
& + \eta D G_1G_2 (m_{br_1} + n_{sp})\big(1+2\eta  m_{bd} \big)h_{rr}h_{rd}  \nonumber \\
&+ \eta D G_2 (m_{br_2} + n_{sp})\big(1+ 2\eta m_{bd}\big)h_{rd} \nonumber \\
&+ D\big(\eta G_1G_2 (m_{br_1} + n_{sp})\big)^2 h_{rr}^2 h_{rd}^2    \nonumber \\
&+ D\big(\eta G_2 (m_{br_2} + n_{sp})\big)^2 h_{rd}^2 \nonumber \\
&+ 2\eta^2 G_1G_2^2 (m_{br_1} + n_{sp}) (m_{br_2} + n_{sp})h_{rr}h_{rd}^2 \nonumber \\
&+ 2\eta^2 G_1^2G_2^2 m_s(m_{br_1} + n_{sp}) h_{sr} h_{rr}^2h_{rd}^2 \nonumber \\
&+ 2\eta^2 G_1  G_2^2 m_s(m_{br_2} + n_{sp}) h_{sr} h_{rr}  h_{rd}^2     \nonumber \\
&+ \eta D \big( m_{bd} +  m_{bd}^2\big). 
\end{align}
Because of the large number of photons after optical amplification, the mean and the variance of the received photon count at the destination can be approximated closely by a Gaussian \cite{razavi_2005}. 
%This simplifies derivation of closed-form analytical expressions for evaluating the system performance.
Moreover, the thermal noise is generated independently from the received optical signal and has a zero-mean Gaussian distribution with variance due to the resistive element
\begin{align}
\label{therm_1}
\sigma^2_{th} = \frac{2 K_B T_R T_s}{R_L e^2},
\end{align}
where $K_B$ is the Boltzmann constant, $T_R$ and $R_L$ are the receiver’s equivalent temperature and load, and $e$ denotes the electron charge.
While dark current noise can practically be neglected, the photo-electron count at the destination can accurately be modeled by a Gaussian distribution with mean $m_d = m'_d$ and variance
\begin{align}
\label{sigma2d}
\sigma_d^2 = \sigma'^2_d + \sigma_{th}^2.
\end{align}

Finally, the electrical SNR at destination is expressed as 
\begin{align}
\label{pj_1}
\gamma = \left(\eta G_1G_2 m_s h_{sr} h_{rr} h_{rd}\right)^2/\sigma_d^2.
\end{align}

As we observe from \eqref{lmw2} and \eqref{sigma2d}, the variance of noise at the destination is very complex. 
In order to simplify, in the sequel, we propose two approximations over noise variance and then, in the Section III we investigate the accuracy of these approximations.
First, we consider the practical case where the amount of background power is low. At this condition, the noise variance at the destination can be approximated as 
 \begin{align}
\label{lmw6}
\sigma^2_d\simeq&~2\eta^2  G_1^2G_2^2 m_s(m_{br_1} + n_{sp}) h_{sr} h_{rr}^2h_{rd}^2 \nonumber \\
&+ 2\eta^2  G_1  G_2^2 m_s(m_{br_2} + n_{sp}) h_{sr} h_{rr}  h_{rd}^2   + \sigma_{th}^2. 
\end{align}
To study the effect thermal noise has on the performance of the considered dual hop FSO system, we assume that the SNR at the destination is limited by thermal noise. According this assumption, noise variance at the destination is simplified as 
\begin{align}
\label{h1}
\sigma_d^2 \simeq  \sigma^2_{th}.
\end{align}
%%%%%%%%%%%%%%%%%%%%%%%%%%%%%%%%%%%%%
%\subsection{Performance Analysis}
%%%%%%%%%%%%%%%%%%%%%%%%%%%%%%%%%%%%%

%where $h_{th}= \sqrt{\sigma_n^2\Upsilon_{th}}/R P_t$, and $f_h$ and $f_\Upsilon$ denote the distribution of $h$ and $\Upsilon$, respectively.
%%%%%%%%%%%%%%%%%%%%%%%%%%%%%%%%%%%%%%%%%%%%%
%%%%%%%%%%%%%%%%%%%%%%%%%%%%%%%%%%%%%%%%%%%%%
\section{Numerical Results and Discussions}
%%%%%%%%%%%%%%%%%%%%%%%%%%%%%%%%%%%%%%%%%%%%%
%%%%%%%%%%%%%%%%%%%%%%%%%%%%%%%%%%%%%%%%%%%%%
%%%%%%%%%%%%%%%%%%%%%%%%%%%%%%%%%%%%%%%%%%%%%%%%%%%%%%%%%%%%%%%%
%%%%%%%%%%%%%%%%%%%%%%%%%%%%%%%%%%%%%%%%%%%%%%%%%%%%%%%%%%%%%%%%
We now present numerical and analytical results to evaluate the performance of the considered triple-hop FSO system over log-normal turbulence channel.
Different system parameters used throughout performance evaluation are provided in Table \ref{par2}. 
To facilitate numerical analysis, we consider a similar condition for SR, RR and RD links, i.e., parameters such as visibility $\mathcal{V}$, refractive index structure constant $C_n^2$, background power $P_{br1}=P_{br1}=P_{bd}=P_b$, and divergence angle $\theta_{div}$, are set equal in both channels.

In this paper, we analyze the performance of considered system in term of ergodic or average capacity. 
As we observe from \eqref{pj_1}, electrical SNR  at the destination is a function of $h_{sr}$, $h_{rr}$ and $h_{rd}$. Hence, for considered system, ergodic capacity can be obtained as
\begin{align}
\label{ko2}
\mathcal{C}_{\rm erg}=& \int_0^\infty \!\!\int_0^\infty\!\! \int_0^\infty \log\Big(1+\gamma(h_{sr},h_{rr}, h_{rd}) \Big)dh_{sr}dh_{rr}dh_{rd}.
\end{align}

%%%%%%%%%%%%%%%%%%%%%%%%%%%%%%%%
%%%%%%%%%%%%%%%%%%%%%%%%%%%%%%%%
\begin{table}
	\caption{System Parameters Used for Simulations} 
	\centering
	\begin{tabular}{l c c}
		\hline\hline \\[-1.2ex]
		{\bf Description} & {\bf Parameter} & {\bf Setting} \\ [.5ex] 
		\hline\hline \\[-1.2ex]
		Wavelength                      &$ \lambda $      &$ 1550$ nm   \\[1ex] 
		Source-to-destination length    &$d_{sd}$& 5 km \\[1ex]
		Visibility                      &$\mathcal{V}$& 1.5 km    \\[1ex]
		Refractive index                &$C_n^2$& $10^{-15}$   \\[1ex]  
		Aperture area                   & $ A_a $   &  $\pi (0.1)^2$  \\[1ex] 
		Divergence angle                & $ \theta_{div} $    &$ 1 $ mrad  \\[1ex] 
		Plank\textquotesingle s Constant&$ h_{p} $        &$ 6.6 \times 10^{-34} $  \\[1ex]              
		Boltzmann\textquotesingle s Constant&$ K_{B} $    & $ 1.38 \times 10^{-23} $ \\[1ex]                
		Receiver Load                   &$ R_{L} $        & $ 100~\Omega $ \\[1ex]                
		Receiver Temperature            &$ T_{R} $        & $ 300\degree~K$  \\[1ex] 
		Transfer Rate                   & $R_s$           & 2 Gbit/s     \\[1ex]               
		Symbol Duration                 &$ T_{s} $        & $1/R_s$ s\\[1ex]  
		Spontaneous emission parameter  &$ n_s $          &  1\\[1ex] 
		DoF                             &$ D $            &  100\\[1ex] 
		
		Quantum efficiency              & $ \eta $         & 0.8 \\[1ex]
		Average transmit power of source        & $ P_s $         & 5 dBm W \\[1ex]
		Average transmit power of relays         & $ P_{r_1}\&P_{r_2} $         & 5 dBm W \\[1ex]
		\hline\hline 
	\end{tabular}
	\label{par2}
\end{table}
%%%%%%%%%%%%%%%%%%%%%%%%%%%%%%%%%%%%%%%%%%%%%%%%%%
%%%%%%%%%%%%%%%%%%%%%%%%%%%%%%%%%%%%%%%%%%%%%%%%%%
%%%%%%%%%%%%%%%%%%%%%%%%%%%%%%%
%%%%%%%%%%%%%%%%%%%%%%%%%%%%%%%
\begin{figure}
	\begin{center}
		\includegraphics[width=3.3 in ]{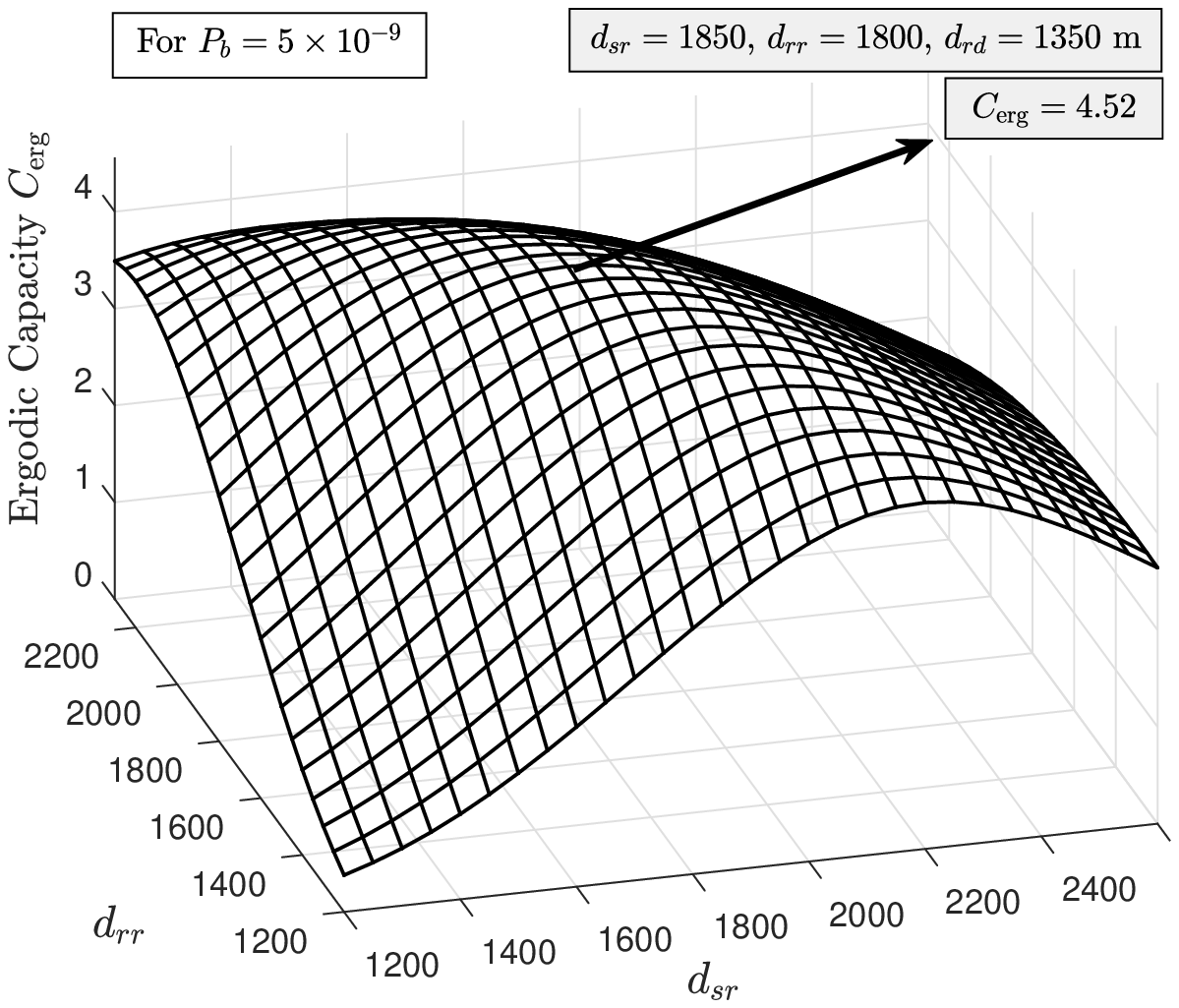}
		\caption{ Average capacity of FSO relaying system versus SR link length $d_{sr}$, and  RR link length $d_{rr}$ for a fixed SD link length $d_{sd}=5$ km and  $P_b=5\times10^{-9}$.}
		\label{1}
	\end{center}
\end{figure}
%%%%%%%%%%%%%%%%%%%%%%%%%%%%%%%
%%%%%%%%%%%%%%%%%%%%%%%%%%%%%%%%
%
%%%%%%%%%%%%%%%%%%%%%%%%%%%%%%%
%%%%%%%%%%%%%%%%%%%%%%%%%%%%%%%
\begin{figure}
	\begin{center}
		\includegraphics[width=3.3 in ]{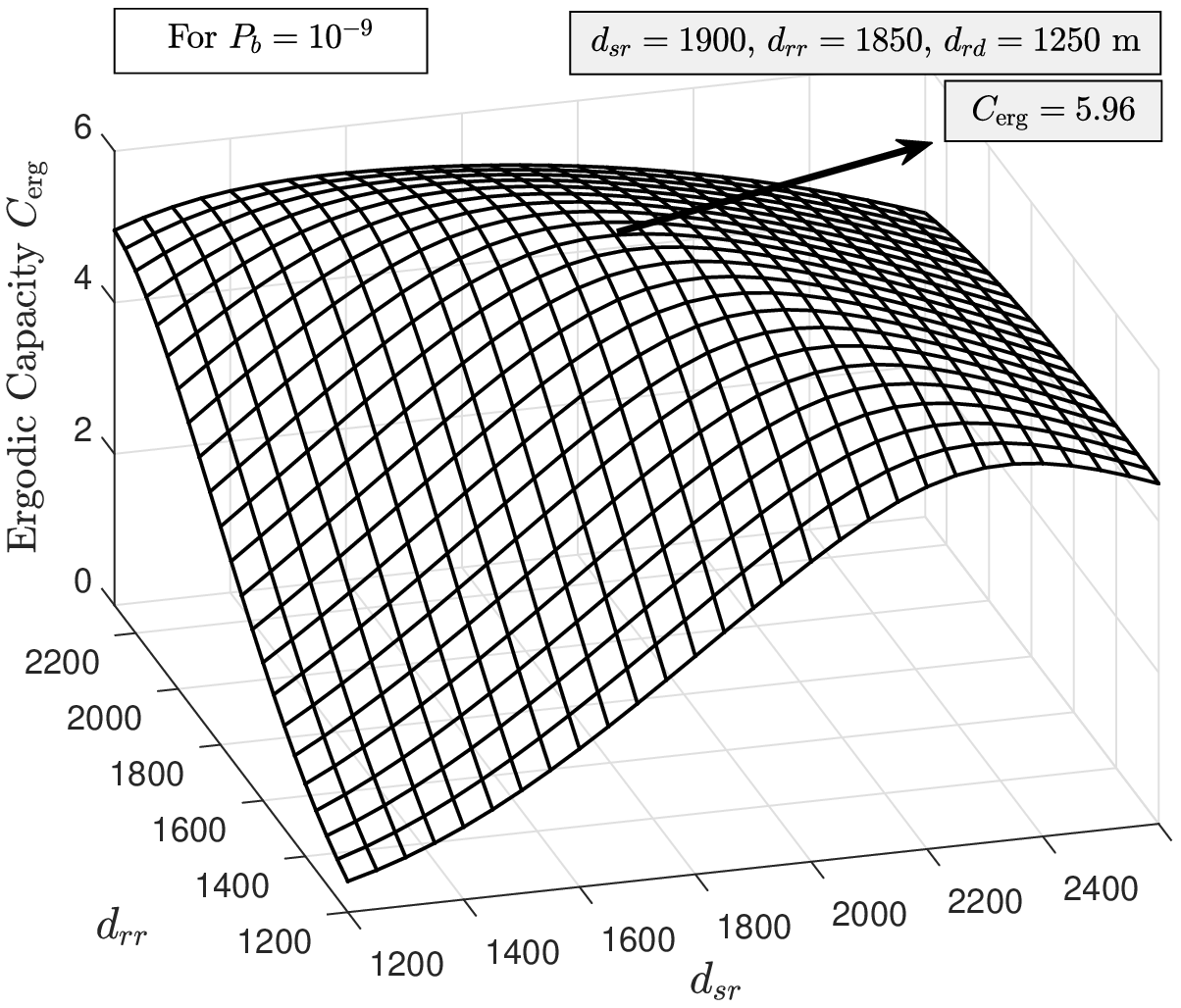}
		\caption{ Average capacity of FSO relaying system versus SR link length $d_{sr}$, and  RR link length $d_{rr}$ for a fixed SD link length $d_{sd}=5$ km and  $P_b=10^{-9}$.}
		\label{2}
	\end{center}
\end{figure}
%%%%%%%%%%%%%%%%%%%%%%%%%%%%%%%
%%%%%%%%%%%%%%%%%%%%%%%%%%%%%%%
%
%
%%%%%%%%%%%%%%%%%%%%%%%%%%%%%%%
%%%%%%%%%%%%%%%%%%%%%%%%%%%%%%%
\begin{figure}
	\begin{center}
		\includegraphics[width=3.3 in ]{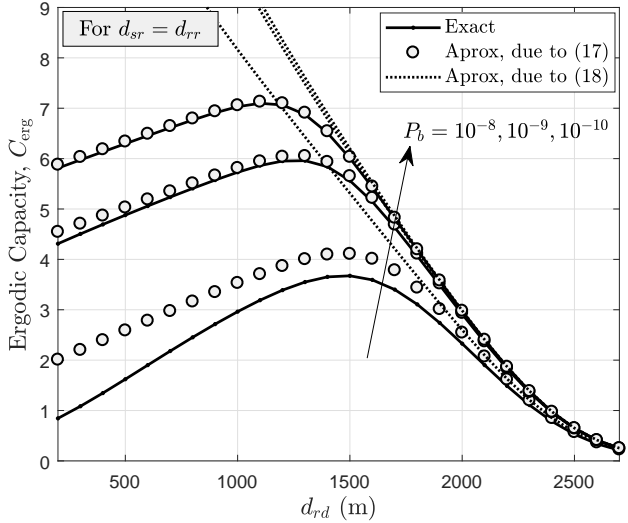}
		\caption{ Average capacity of FSO relaying system versus RD link length $d_{rd}$ for a fixed SD link length $d_{sd}=5$ km and different values of  $P_b$.}
		\label{3}
	\end{center}
\end{figure}
%%%%%%%%%%%%%%%%%%%%%%%%%%%%%%%
%%%%%%%%%%%%%%%%%%%%%%%%%%%%%%%%

To demonstrate the effects of optimum relay location on the performance of the considered triple-hop FSO system, in Figs. \ref{1} and \ref{2}, we have plotted the ergodic capacity versus SR link length $d_{sr}$, and  RR link length $d_{rr}$ for a fixed SD link length $d_{sd}=5$ km. 
Figure \ref{1} is depicted for $P_b=5\times10^{-9}$ and Fig. \ref{2} is depicted for $P_b=10^{-9}$.
There are three important observations which can be drawn from these figures.
First, the performance  strongly depends on the relay locations and as we observe, by changing relay locations the ergodic capacity changes, significantly. 
Second, the optimum location of relays (the location with higher ergodic capacity) depends on the background noise level.
For instance, for $P_b=5\times10^{-9}$, the optimum location of relays are achieved at $d_{sr}=1850$, $d_{rr}=1800$ and $d_{rd}=1350$ m and for $P_b=10^{-9}$, the optimum location of relays are achieved at $d_{sr}=1900$, $d_{rr}=1850$ and $d_{rd}=1250$ m.
Third, at the optimum location of relays, SR link length is approximately equal to RR link length and they are bigger than that RD link length.

Now we want to investigate the validity of approximations on \eqref{lmw2} which are characterized by  \eqref{lmw6} and \eqref{h1}.
To this aim, in Fig. \ref{3}, we have plotted the ergodic capacity versus $d_{rd}$  for different values of $P_b$ where $d_{sr}=d_{rr}$.
In addition to the conclusions drawn form the simulation results proposed in the previous section, from Fig. \ref{3} we can draw the  conclusions that at low values of $P_b$,  our proposed approximation in \eqref{lmw2} for the noise variance at the destination is valid for all of possible relay locations, while the thermal noise limited approximation proposed in \eqref{h1} is valid for locations close to the source.

%%%%%%%%%%%%%%%%%%%%%%%%%%%%%%%%%%%%%%%%%%%%%%%%%%%
%%%%%%%%%%%%%%%%%%%%%%%%%%%%%%%%%%%%%%%%%%%%%%%%%%%
%\vspace{-3mm}
\section{Conclusion}
%%%%%%%%%%%%%%%%%%%%%%%%%%%%%%%%%%%%%%%%%%%%%%%%%%%
%%%%%%%%%%%%%%%%%%%%%%%%%%%%%%%%%%%%%%%%%%%%%%%%%%%
In this paper, we proposed a comprehensive signal model of triple-hop all-optical relaying FSO systems in the presence of all main noise sources including background, thermal and  ASE noise by considering at the same time the effect of the DoF.
Using full CSI relaying, we derived the exact expressions for the noise variance at the destination.
Then, in order to simplify the analytical expressions of full CSI relaying, we also proposed and investigated the validity of different approximations over noise variance at the destination.
Numerical results were provided in term of ergodic capacity and they confirmed that for low values of the background power, the proposed approximation regarding SNR at the destination is valid for all of possible relay locations. For thermal noise limited, it was shown that the considered approximation is valid when the relay is close to the source.

%%%%%%%%%%%%%%%%%%%%%%%%%%%%
%%%%%%%%%%%%%%%%%%%%%%%%%%%%
%\bibliographystyle{IEEEtran}
%\balance
%\bibliography{IEEEabrv,myref}
%%%%%%%%%%%%%%%%%%%%%%%%%%
\balance 
%%%%%%%%%%%%%%%%%%%%%%%%%%%%%%%%%%%%%%%%%%%%%%%%%%%%%%%%%%%%%%%%

%%%%%%%%%%%%%%%%%%%%%%%%%%%%%%%%%%%%%%%%%%%%%%%%%%%%%%%%%%%
%\balance 

\end{document}